\documentclass[aps,prd,onecolumn,amssymb]{revtex4}
\usepackage{graphicx,bm,color}
\usepackage{amsmath}
\usepackage{amsfonts}
\usepackage[latin1]{inputenc}
\usepackage[T1]{fontenc}
\usepackage{subfigure}
\usepackage{fancyhdr}
\usepackage{mathrsfs}
\usepackage{color,psfrag}

\def\({\left(}
\def\){\right)}

\newcommand{\beq}{\begin{eqnarray}}
\newcommand{\eeq}{\end{eqnarray}}

\def\bce{\begin{center}}
\def\ece{\end{center}}

\newcommand{\be}{\begin{equation}}
\newcommand{\ee}{\end{equation}}
\newcommand{\bea}{\begin{eqnarray}}
\newcommand{\eea}{\end{eqnarray}}
\newcommand{\beaa}{\begin{eqnarray*}}
\newcommand{\eeaa}{\end{eqnarray*}}

\newcommand{\e}{\mathrm{e}}

\begin{document}

\title{Multiple $\Lambda$CDM cosmology with
string landscape features \\ and future singularities}
\author{E.~Elizalde$^{a,}$\footnote{E-mail: elizalde@ieec.uab.es,
elizalde@math.mit.edu}, A.~N.~Makarenko$^{b,}$\footnote{E-mail:
andre@tspu.edu.ru},
S.~Nojiri$^{c,d,}$\footnote{E-mail: nojiri@phys.nagoya-u.ac.jp},
V.~V.~Obukhov$^{b,}$\footnote{E-mail:
obukhov@tspu.edu.ru} and S.~D.~Odintsov$^{a,e,b,}$\footnote{E-mail:
odintsov@ieec.uab.es}}
\affiliation{$^a$Consejo Superior de Investigaciones Cient\'{\i}ficas, ICE/CSIC-IEEC \\
Campus UAB, Facultat de Ci\`{e}ncies, Torre C5-Parell-2a pl, E-08193
Bellaterra (Barcelona) Spain \\
$^b$Department of Theoretical Physics, Tomsk State Pedagogical University,
Tomsk, 634041 Russia \\
$^c$Department of Physics, Nagoya University, Nagoya
464-8602, Japan \\
$^d$Kobayashi-Maskawa Institute for the Origin of Particles and
the Universe, Nagoya University, Nagoya 464-8602, Japan \\
$^e$Instituci\'{o} Catalana de Recerca i Estudis Avan\c{c}ats (ICREA) }

\begin{abstract}
Multiple $\Lambda$CDM cosmology is studied in a way that is formally a
classical analog of the Casimir effect. Such cosmology corresponds to a
time-dependent dark fluid model or, alternatively, to its scalar field
presentation, and it motivated by the string landscape picture.
The future evolution of the several dark energy models constructed within
the scheme is carefully investigated. It turns out to be almost always
possible to choose the parameters in the models so that they match the
most recent and accurate astronomical values. To this end, several
universes are presented which mimick (multiple) $\Lambda$CDM cosmology but
exhibit Little Rip, asymptotically de Sitter, or Type I, II, III, and IV
finite-time singularity behavior in the far future, with disintegration
of all bound objects in the cases of Big Rip, Little Rip and Pseudo-Rip
cosmologies.
\end{abstract}


\maketitle

\section{Introduction}

Astronomical observations indicate that our Universe is currently in an
accelerated phase \cite{Dat}. This acceleration in the expansion rate of
the observable cosmos is usually explained by introducing the so-called
dark energy (for a recent review, see \cite{review}).
In the most common models considered in the
literature, dark energy comes from an ideal fluid with a specific
equation of state (EoS) often exhibiting rather strange properties, as
a negative pressure and/or a negative entropy, and also the fact that its
action was invisible in the early universe while it is dominant in our
epoch, etc.
According to the latest observational data, dark energy currently accounts
for some 73\% of the total mass-energy of the universe (see, for example,
Ref.~\cite{Kowalski}).

In an attempt at saving General Relativity and to explain the cosmic
acceleration, at the same time, one is led to conjecture some exotic dark
fluids (although some other variants are still being considered, see e.g.
\cite{vari1}). Actually, General Relativity with an ideal fluid can be
rewritten, in an equivalent way, as some modified gravity. Also, the
introduction of a fluid with a complicated equation of state is to be seen
as a phenomenological approach, since no explanation for the origin of
such dark fluid is usually available. However, the interesting possibility
that the dark fluid origin could be related with some fundamental theory,
as string theory, opens new possibilities, through the sequence: string or
M-theory is approximated by modified (super)gravity, which is finally
observed as General Relativity with an exotic dark fluid. If such
conjecture would be (even partially) true, one might expect that some
string-related phenomena could be traceable in our dark energy universe.
One celebrated stringy effect possibly related with the early universe
comes from the string landscape (see, for instance, \cite{land}), which
may lead to some observational consequences (see, e.g., \cite{mar}), since
it could be responsible for the actual discrete mass spectrum of scalar
and spinorial equations \cite{igor}.

The equation of state (EoS) parameter $w_\mathrm{D}$ for dark energy is
negative:
\begin{equation}
w_\mathrm{D}=p_\mathrm{D}/\rho_\mathrm{D}<0\, ,
\end{equation}
where $\rho_\mathrm{D}$ is the dark energy density and $p_\mathrm{D}$ the
pressure. Although astrophysical observations favor the standard
$\Lambda$CDM cosmology, the uncertainties in the determination of the EoS
dark energy parameter $w$ are still too large, namely $w=-1.04^{+0.09}_{-0.10}$,
to be able to determine, without doubt, which of the three cases:
$w < -1$, $w = -1$, and $w >-1$ is the one actually realized in our universe
\cite{PDP,Amman}.

The phantom dark energy case $w < -1$, is most interesting but poorly
understood theoretically. A phantom field violates all four energy
conditions, and it is unstable
from the quantum field theoretical viewpoint, although it still could be
stable in classical cosmology. Some observations hint towards a possible
crossover of the phantom divide in the near past or in the near future. A
very unpleasant property of phantom dark energy is the appearance of a Big
Rip future singularity \cite{kam}, where the scale factor becomes infinite at
finite time in the future.
A less dangerous future singularity caused by phantom
or quintessence dark energy is the sudden (Type II) singularity \cite{barrow}
where the scale factor is finite at Rip time.
Closer examination shows, however, that the condition $w<-1$ is not
sufficient for a singularity occurrence.
First of all, a transient phantom cosmology is quite possible.
Moreover, one can easily construct models where $w$ asymptotically
tends to $-1$ and such that the energy density increases with time, or
remains constant, but there is no finite-time future singularity, what
was extensively studied in Refs.~\cite{kam,Nojiri-3,barrow,Stefanic,Sahni:2002dx}
(for a review, see \cite{review}, and for their classification, \cite{Nojiri-3}).
A clear case is when the Hubble rate tends to a constant (a cosmological
constant or asymptotically de Sitter space), which may also correspond to
a pseudo-Rip situation \cite{Frampton-3}. Also to be noted is the
so-called Little Rip cosmology \cite{Frampton-2}, where the Hubble rate
tends to infinity in the infinite future (for further details, see
\cite{Frampton-3,LR}).
The key point is that if $w$ approaches $-1$ quickly enough,
then it is possible to have a model in which the time required for the
singularity to appear is infinite, so that the singularity never forms in practice.
Nevertheless, it can be shown that even in this case the disintegration of bound structures
takes place, in a way similar to the Big Rip phenomenon. Such models are known as
Little Rip and they have both a fluid and a scalar field description \cite{Frampton-2,As1}.

In the present paper we investigate a dark fluid model with a time-dependent
EoS which can be considered as simple classical analog of the string
landscape \cite{land_odin}.
The Casimir effect may lead to a similar picture. Some vacuum states appear
which can be implemented with the help of the landscape. Moreover, we will
study multiple $\Lambda$CDM cosmology as a classical analog of the Casimir
effect (for a review see \cite{Caz1}). This cosmology is also motivated by
the string landscape picture. We demonstrate that such multiple $\Lambda$CDM
cosmology may lead to various types of future universe, not only the asymptotically
de Sitter one, but also to Little Rip cosmology and a finite-time future
singularity, of any of the four known types \cite{Nojiri-3}. The equivalent
description of multiple $\Lambda$CDM cosmology in terms of scalar theory
is also further developed.

\section{Ideal fluid  leading to multiple $\Lambda$CDM cosmology}

Let us study the specific model of an ideal fluid which leads to a multiple
$\Lambda$CDM cosmology. The corresponding FRW equations are
\begin{equation}
\label{k3}
\frac{3}{\kappa^2}H^2 = \rho\, ,\quad
  - \frac{2}{\kappa^2}\dot H= p + \rho\, .
\end{equation}
Here $\rho$ is the energy density and $p$ the pressure.
Instead of (\ref{k3}), one can include the cosmological constant from
gravity:
\begin{equation}
\label{k3B}
\frac{3}{\kappa^2}H^2 = \rho+ \frac{\Lambda}{\kappa^2}\, ,\quad
  - \frac{2}{\kappa^2}\dot H= p + \rho\, .
\end{equation}
We can, however, redefine $\rho$ and $p$ in order to absorb the contribution
coming from the cosmological constant, namely,
\begin{equation}
\label{k3C}
\rho\to \rho - \frac{\Lambda}{\kappa^2}\, ,\quad
p\to p + \frac{\Lambda}{\kappa^2}\, .
\end{equation}
With the redefinition (\ref{k3C}), we re-obtain (\ref{k3}).
Hence, it is enough to consider only the dark fluid in the FRW equation.

If $\rho$ and $p$ are given in terms of the function $f(q)$, with a
parameter $q$ given by (compare with the similar Ansatz in
\cite{land_odin,grg})
\begin{equation}
\label{kk1}
\rho = \frac{3}{\kappa^2}f(q)^2\, ,\quad
p=-\frac{3}{\kappa^2}f(q)^2 - \frac{2}{\kappa^2}f'(q)\, ,
\end{equation}
the following solution of Eq.~(\ref{k3}) is found
\begin{equation}
\label{ML1}
H=f(t)\, .
\end{equation}
Note that the origin of time can be chosen arbitrarily.
In (\ref{ML1}), $t=q$ but one may choose $t=q+t_0$ with an
arbitrary constant $t_0$.
This shows that, besides the solution (\ref{ML1}),
$H=f(t-t_0)$ can also be a solution.

If we delete $q$ in (\ref{kk1}), we obtain a general equation of state
(EoS):
\begin{equation}
\label{ML2}
F(\rho,p)=0\, .
\end{equation}
In the case that $f'(q)=0$ has a solution $q=q_0$, then there is a solution
in which $H$ is a constant:
\begin{equation}
\label{ML3}
H=H_0\equiv f(q_0)\, ,
\end{equation}
where $\rho=-p$, what corresponds to an effective cosmological constant.
Then, if there is more than one solution satisfying $f'(q)=0$, as $q=q_n$,
$n=0,1,2, \cdots$, the theory could effectively admit several different
cosmological constants, namely
\begin{equation}
\label{ML3b}
H=f(q_n)\, ,\quad \Lambda_n=3f(q_n)^2\, .
\end{equation}
Note that, indeed, solutions (\ref{ML3b}) corresponding to
different cosmological constants can
exist simultaneously, which shows an interesting analogy with the
cosmological landscape situation in string/M theory.
Let us assume that, in fact, there is a solution corresponding to $q_n$.
By perturbing this solution it may transit to
another one, say $q_{n+1}$. The transition period will be
proportional to $T_{n,n+1}=q_{n+1} - q_n$. This also hints to the
possibility of occurrence of several $\Lambda$CDM phases in our observable
universe.

\section{Example 1: Non-periodic behavior of the dark fluid}

Consider the simplest case with two values of the cosmological constant
\begin{equation} \label{ex1}
\dot{H}=q (\Lambda_1-t)(\Lambda_2-t)(1+\beta t)^\gamma\, ,
\end{equation}
where $q$, $\Lambda_1$, $\Lambda_2$, and $\gamma$ are constants.
In this case the Hubble parameter takes the form ($\gamma \ne -1,\,-2,\,
-3$)
\begin{eqnarray}
H&=&q\frac{(1+\beta t)^{1+\gamma}}{\beta^3 (1+\gamma) (2+\gamma)
(3+\gamma)} \\
&& \times \left(2+\beta \left((3+\gamma) (\Lambda_2+\Lambda_1
(1+\beta \Lambda_2 (2+\gamma)))-(1+\gamma) (2+\beta (\Lambda_1+\Lambda_2)
(3+\gamma)) t+\beta (1+\gamma) (2+\gamma) t^2\right)\right)\, .\nonumber
\end{eqnarray}
It is easy to find the form of the scale factor ($\gamma \ne -4$)
\begin{equation}
a(t)=a_0 \e^{q\frac{(1+\beta t)^{2+\gamma} \left(6+\beta \left((4+\gamma)
(2 \Lambda_2+\Lambda_1 (2+\beta\Lambda_2 (3+\gamma)))-(1+\gamma) (4+\beta
(\Lambda_1+\Lambda_2) (4+\gamma)) t+\beta (1+\gamma) (2+\gamma)
t^2\right)\right)}{\beta^4 (1+\gamma) (2+\gamma) (3+\gamma) (4+\gamma)}}\, .
\end{equation}
By choosing different values for the constants, the model will have
different behaviors. Thus, if $\gamma>0$ then, for large values of time,
the Hubble constant will be proportional to $t^{3+\gamma}$. If $\gamma<-4$
we have that $H \sim t^{3-|\gamma|}$.  In addition, the constant $\beta$
can be either positive or negative. In the second case we obtain a
singularity in the future:
$a$ and $ \rho$ go to infinity at finite time.

Suppose now that $\Lambda_1 = 0.1$ and $\Lambda_2=13.6$, at these points
where we have an effective cosmological constant. The current value of
the Hubble constant is known to be $H_0^{-1}=13.6$ Gyr. So one can find
the value of the constant $q$.
Now, choosing the value of $\gamma$, we can find $\beta$, the jerk
parameter $j_{0}$ having been used in accordance with the observations.
The deceleration parameter $q_0$ is $-1$, since at the current time we
have a model with an effective cosmological constant.
The calculated values of both the deceleration parameter $q_0$ and the
jerk parameter $j_0$ can be found in Ref.~\cite{Rapetti}:
$q_{0}=-0.81\pm0.14$ and $j_{0}=2.16^{+0.81}_{-0.76}$ (from
type Ia supernovae and X-ray cluster gas mass fraction measurements).

We choose, as an example of two parameter values for gamma: $\gamma=12$ and
$\gamma=-5$.
In the first case, in order for the jerk parameter to be in the
permissible region, it is necessary that the parameter $\beta$ be in the
range  $0.00433706 < \beta < 0.00660997$.  In the second case, we have
that $-0.0228368<\beta<-0.0164559$.
Thus, for this choice of constants, we have the following values of the
cosmological parameters:
\[
j_{0}=2.16^{+0.81}_{-0.76}\, ,\quad q_0=-1\, ,\quad H_0^{-1}=13.6\, \mathrm{Gyr}\, ,
\quad w=-1\, .
\]
Assume that, at present, our model is approaching, or has
already passed, the point corresponding to an effective cosmological
constant. Let us set $\gamma=12$, then we can bring the model to the
desired set up $q_0$ value.  For $\Lambda_2<t_0$  one cannot choose the
parameter $\beta$ in order to do the same.

Assume that $\Lambda_2=14$ ($t_0=13.6$). Then, the parameter $\beta$ has
to take values in the range: $0.00637252351<\beta<0.006847247$.
Thus, for this choice of constant, we have the following values for the
cosmological parameters:
\[
2.452< j_{0}<2.97\, ,\quad -0.95>q_0>-0.932\, ,\quad
H_0^{-1}=13.6\, \mathrm{Gyr}\, ,\quad -0.967<w<-0.955\, .
\]
As we see, in this case $w>-1$.

Suppose now that $\gamma=-5$ and $\Lambda_2=14$ ($t_0=13.6$), then
$-0.0222>\beta>-0.02364$, and we have the following
cosmological parameters:
\[
2.3915< j_{0}<2.45178\, ,\quad -0.95>q_0>-0.92868\, ,\quad
H_0^{-1}=13.6\, \mathrm{Gyr}\, ,\quad -1.06942<w<-1.0492\, .
\]
Note that in this case $w<-1$ and we will have a Big Rip future singularity
($\rho$, $p$, $a \to \infty$) for $t$ in the range $42.2836<t<45.043$
(the lifetime of the universe).

\begin{figure}[-!h]
\begin{center}
\begin{minipage}[h]{0.4\linewidth}
\includegraphics[angle=0, width=1\textwidth]{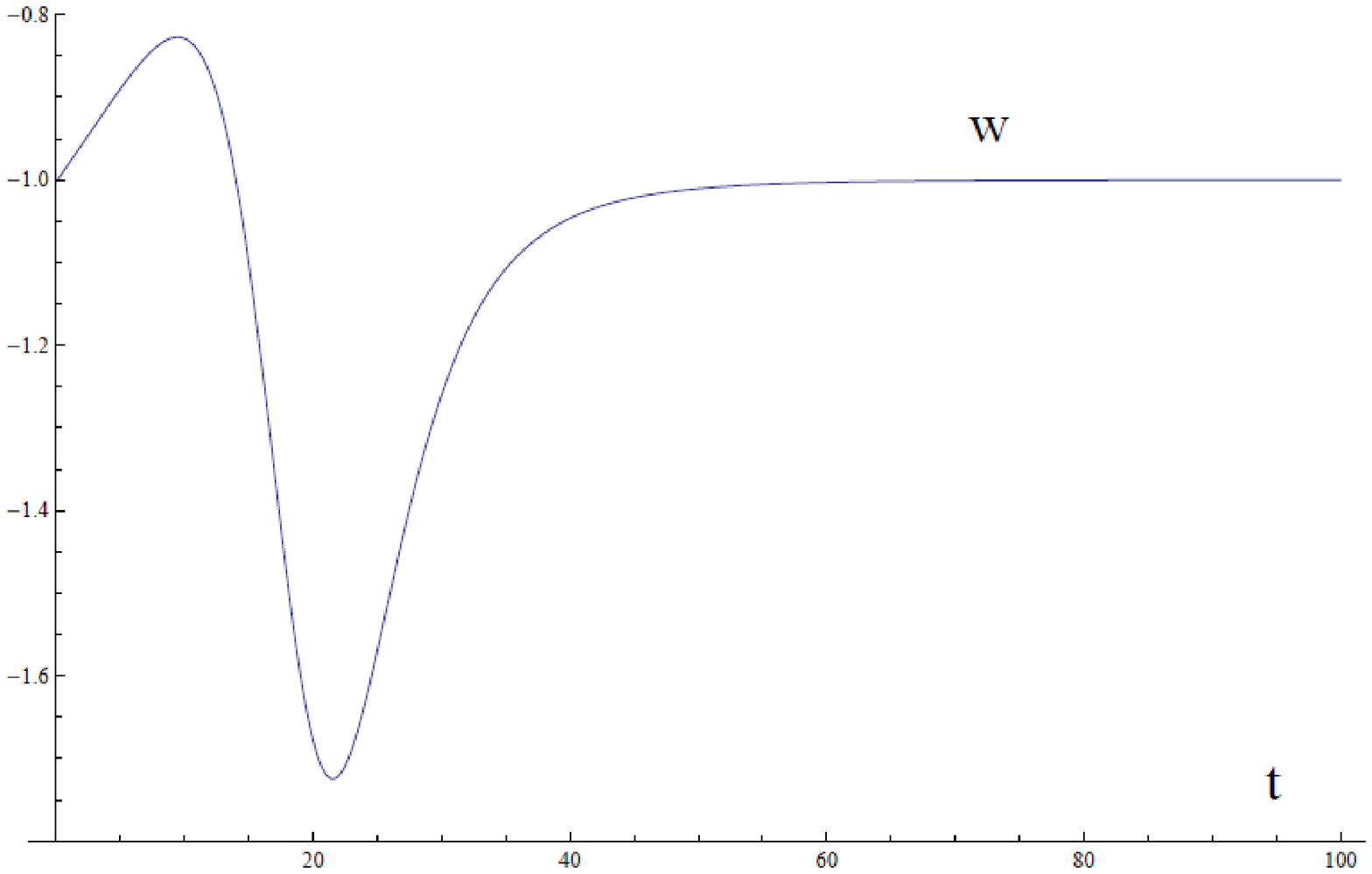}
\caption{\label{Fig1} Plot of $w(t)$ ($0\leq t \leq 100$), for $\Lambda_1=0.1$,
$\Lambda_2=14$, $\beta=0.00637252351$, $\gamma=12$, $q=0.0000184648$.}
\end{minipage}
\hfill
\begin{minipage}[h]{0.4\linewidth}
\includegraphics[angle=0, width=1\textwidth]{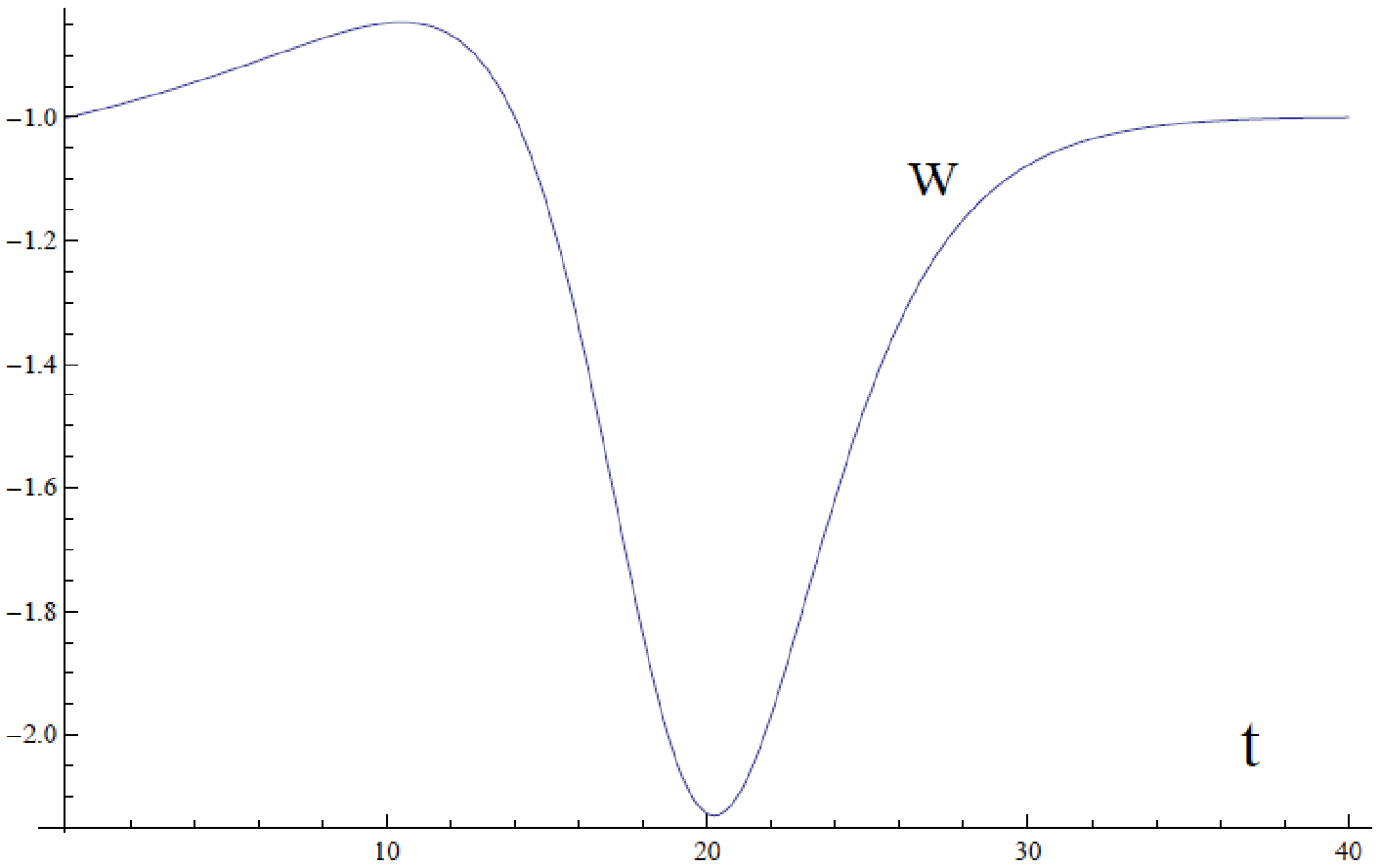}
\caption{\label{Fig2} Plot of $w(t)$ ($0\leq t \leq 100$), for $\Lambda_1=0.1$,
$\Lambda_2=14$, $\beta=-0.023$, $\gamma=-5$, $q=9.329681063413538\times
10^{-6}$.}
\end{minipage}
\end{center}
\end{figure}

Thus, for the chosen model (\ref{ex1}) we have two possible scenarios for
the evolution of the universe:
\begin{enumerate}
\item If $\gamma> 1$, then the behavior of the EoS parameter $w$ is
described in Fig.~\ref{Fig1}, and for $t\to\infty$ we have $w\to-1$ . In this case,
for $t\to\infty$ we obtain that $H\to +\infty$ and we have a ``Little Rip''
\cite{Frampton-2,LR}.
As is known, bound objects in such universe disintegrate.
One can estimate the time required for the solar system
disintegration, the dimensionless internal force being
\begin{equation}
F_\mathrm{iner}=\frac{\ddot{a}}{aH_0^2}\, .
\end{equation}
The Sun-Earth system disintegrates when $F_\mathrm{iner}\sim 10^{23}$ and we
find this time to be $563.58$ Gyr (here $\Lambda_1=0.1$, $\Lambda_2=14$,
$\beta=0.00637252351$, $\gamma=12$, and $q=0.0000184648$).
\item If  $\gamma <-4$ then the behavior of EoS parameter is shown
in Fig.~\ref{Fig2} and we see that there is a singularity in the future at finite
time (a Big Rip singularity), and $w\to -1$. After the singularity the
Hubble constant will tend to zero, and the EoS state parameter  will
increase linearly. The lifetime of the universe that we find for the
following values of the constants: $\Lambda_1=0.1$, $\Lambda_2=14$,
$\beta=-0.023$, and $\gamma=-5$, $q=9.329681063413538 \times 10^{-6}$, is
$42.28<t<45.04$.
In the same way one construct other examples of future evolution with Type II
or Type III future singularity.
\end{enumerate}

\section{Example 2: Periodic behavior of dark fluid}

\subsection{The example of $\exp(\sin)$ dark fluid}

\begin{figure}[-!h]
\begin{center}
\begin{minipage}[h]{0.4\linewidth}
\includegraphics[angle=0, width=1\textwidth]{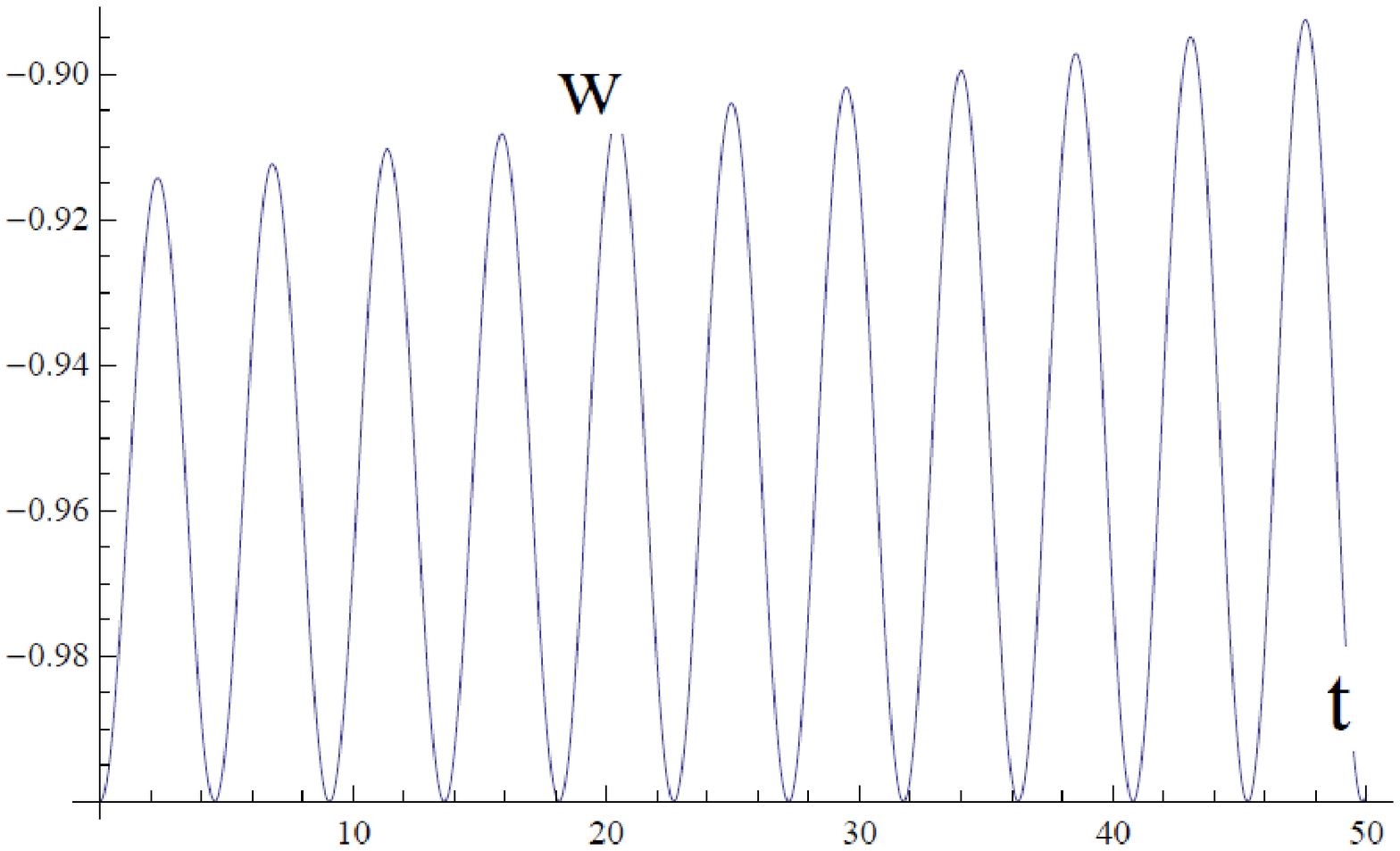}
\caption{\label{Fig3} Plot of $w(t)$ ($0\leq t \leq 50$).}
\end{minipage}
\hfill
\begin{minipage}[h]{0.4\linewidth}
\includegraphics[angle=0, width=1\textwidth]{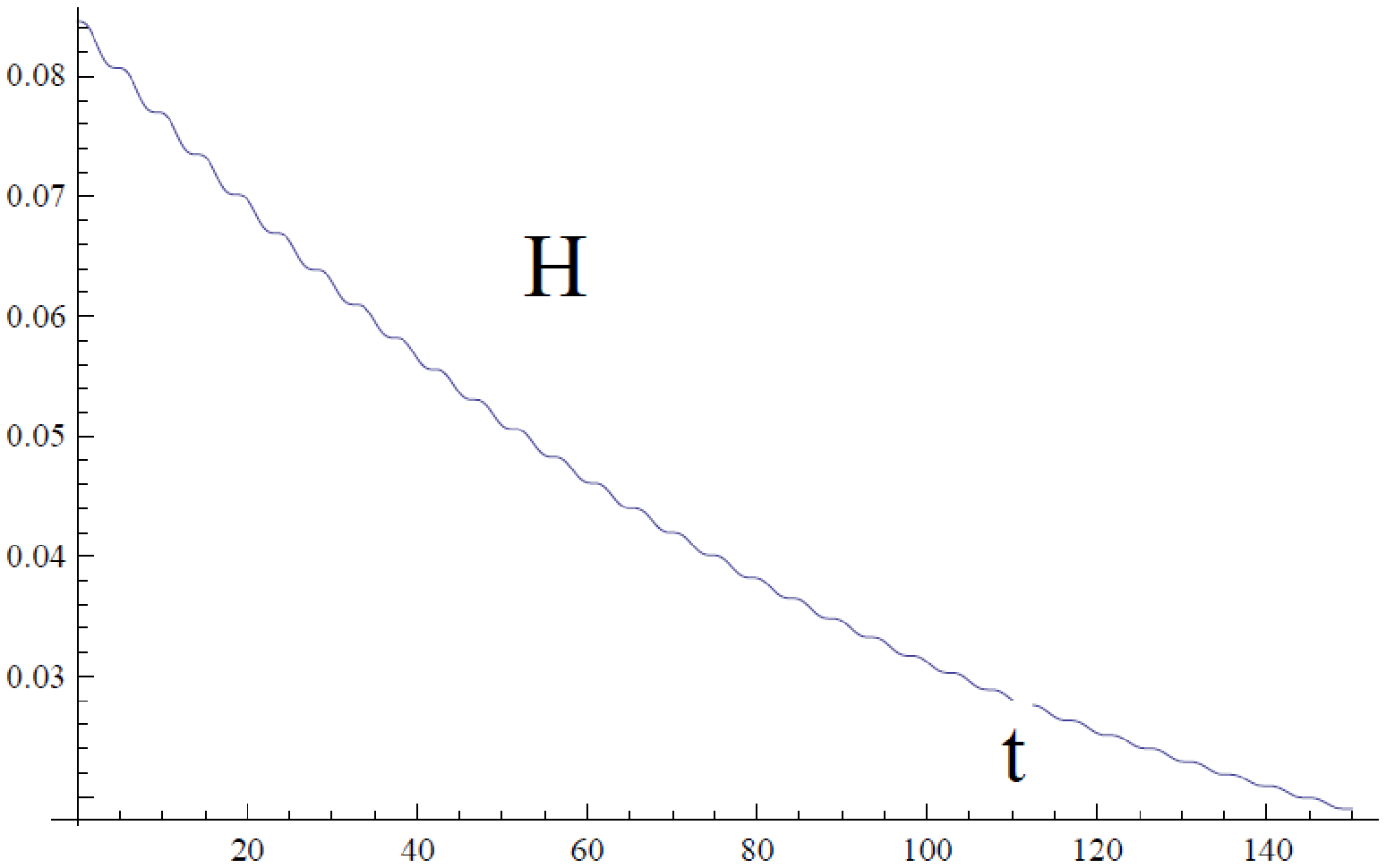}
\caption{\label{Fig4} Plot of $H(t)$ ($0\leq t \leq 150$).}
\end{minipage}
\end{center}
\end{figure}

As a second example, slightly different from the one above,
consider the ideal fluid:
\begin{equation}
\label{MLa1}
f(t)=H=H_0 \e^{ - g \left(t - \frac{1}{\omega} \sin \omega t \right)}\, ,
\end{equation}
which yields
\begin{equation}
\label{MLa2}
f'(t)= - H_0 g \left(1 - \cos \omega t\right)
\e^{ - g \left( 1 - \frac{1}{\omega} \sin \omega t \right)}\, .
\end{equation}
In (\ref{MLa1}), it is assumed that $H_0$, $g$, and $\omega$ are constants.
Therefore, $f'(t)=0$ when $t=\frac{2\pi n}{\omega}$ for integer $n$.
An effective multiple cosmological constant appears as
\begin{equation}
\label{MLa3}
\Lambda_n = 3H_0^2 \e^{ - \frac{4\pi n g}{\omega}}\, .
\end{equation}
Again, $t=2\pi n/\omega$ corresponds to the cosmological constants in
(\ref{MLa1}) and, therefore,
the time-dependent solution could describe the transition between the
cosmological constants, from
the larger to the smaller one. In the limit of $t\to +\infty$ or
$n\to +\infty$, the effective cosmological constant vanishes:
$\lim_{n\to +\infty}  \Lambda_n = 0$.
Now assume that, for $t = 13.6$ Gyr, the Hubble constant is $13.6^{-1}$
Gy$r^{-1}$. We choose the parameters:
\[
\omega=\frac{3 \pi }{7}\, ,\quad g=0.01\, ,\quad H_0=0.084579\, ,
\]
for which we find the following values for the cosmological parameters:
\[
j_0=2.21941\, ,\quad q_0=0.980753\, ,\quad H^{-1}=13.6\, \mbox{\rm Gyr}\, ,\quad
w=-0.987168\, .
\]
The behavior of the Hubble constant is illustrated in Figs.~\ref{Fig3} and
\ref{Fig4}. This
is the ``pseudo-Rip'' case ($H\to H_c=0$, for $t\to\infty$).
In other words, the universe is asymptotically de Sitter one. Nevertheless, due
to the mild phantom behavior of the effective EoS parameter, it
remains the possibility of dissolution of all bound objects sometime in the future.

\subsection{The example $f(\sin)^g$ fluid}

We now consider the following choice for $f(q)$,
\be
\label{p4}
f(q) = H_0 \left( \frac{q}{t_0} - \sin \frac{q}{t_0} \right)\, .
\ee
Here $H_0$ and $t_0$ are positive constants. Then, using (\ref{ML1}),
we get the following solution:
\be
\label{p5}
H = H_0 \left( \frac{t}{t_0} - \sin \frac{t}{t_0} \right)\, .
\ee
Since
\be
\label{p6}
\dot H = \frac{H_0}{t_0} \left( 1 - \cos \frac{t}{t_0} \right)\, ,
\ee
there are de Sitter points, where $\dot H=0$, at $t=2n \pi t_0$, with $n$ an
integer.
When $t\neq 2n \pi t_0$, we find that $\dot H>0$ and, therefore, the
universe is in a phantom phase.
Since $H$ is finite for finite $t$, there is no Big Rip singularity, but
$H$ goes to infinity
when $t$ goes to infinity, that is, a Little Rip occurs.

One can alternatively consider the following $f(q)$,
\be
\label{p7}
f(q) = \frac{H_0}{\left(2N - 1\right) \pi - \left( \frac{q}{t_0}
  - \sin \frac{q}{t_0} \right)}\, ,
\ee
with $N$ a positive integer. Then, $H$ is given by
\be
\label{p8}
H = \frac{H_0}{\left(2N - 1\right) \pi - \left( \frac{t}{t_0} - \sin
\frac{t}{t_0} \right)}\, .
\ee
Again, we find de Sitter points at $t= 2n \pi t_0$, with $n$ integer.
However, when $t\sim \left( 2N - 1 \right) \pi t_0$, instead of the de
Sitter point, we get
\be
\label{p9}
H \sim \frac{H_0}{2 \left( \left(2N - 1\right) \pi - t \right)}\, ,
\ee
which corresponds to a Big Rip singularity.
Therefore, after the de Sitter point $t= 2(N-1) \pi t_0$ or $n=N-1$, there
is a Big Rip singularity
and the universe does never reach the next de Sitter point $t= 2N \pi t_0$.

We may consider more general forms for $f(q)$, as
\be
\label{p10}
f(q) = H_0\left\{ \left(2N - 1\right) \pi - \left( \frac{q}{t_0}
 - \sin \frac{q}{t_0} \right)\right\}^\alpha\, ,
\ee
or
\be
\label{p11}
H = H_0\left\{ \left(2N - 1\right) \pi - \left( \frac{t}{t_0}
 - \sin \frac{t}{t_0} \right)\right\}^\alpha\, ,
\ee
where $\alpha$ is a constant.
Again, we find de Sitter points at $t= 2n \pi t_0$, and for
$t\sim \left( 2N - 1 \right) \pi t_0$,
we find
\be
\label{p12}
H \sim H_0 \left\{2 \left( 2 N \pi - t \right)\right\}^\alpha\, ,
\ee
which corresponds to a Type I (Big Rip) singularity, when $\alpha\leq -1$, to
a Type II one, when $0<\alpha<1$, to one of Type III, when $-1 < \alpha < 0$,
and of Type IV, when $\alpha>1$ and $\alpha$ is not an integer. Already for
the simple model above, the last de Sitter point before the Big Rip
singularity appears at $t= 2(N-1) \pi t_0$.

For the above examples it is not easy at all to write down the EoS explicitly,
by deleting $q$ in (\ref{kk1}), since the EoS obtained often becomes a
multi-valued function.
We should note, however, that it is easy to construct explicit models
with a phantom scalar field to realize the above examples.

We may investigate the deceleration parameter $q_0$ and the jerk parameter
$j_0$, which are defined as
\be
\label{p13}
q_0 = - \frac{1}{aH^2} \frac{d^2 a}{dt^2} = - 1 - \frac{\dot H}{H^2}
\, , \quad
j_0 = 1 + \frac{3\dot H}{H^2} + \frac{\ddot H}{H^3}\, .
\ee
We now evaluate these quantities at the de Sitter points $t=2n\pi t_0$.
For the model (\ref{p5}), we have $H=2n\pi H_0$ and $\dot H = \ddot H = 0$,
and  we find $q_0=-1$ and $j_0=1$.
For the rather simple models (\ref{p9}) and (\ref{p11}), we have already
quite nice results:
$\dot H = \ddot H = 0$, therefore $q_0=-1$ and $j_0=1$. To wit,
in the case of the $\Lambda$CDM model, these parameters are
$q_0 = -0.58$ and $j_0 = 1$.
When the universe is not at a de Sitter point $t\neq 2\pi t_0$, the
universe is in the phantom phase, where
$\dot H>$, and therefore Eq.~(\ref{p13}) tells us tht $q_0 < -1$.
Of course, we neglect the contribution from matter. If we include it, the
universe could not be
in the phantom phase at present, therefore we should obtain $q_0> -1$.

When we do include matter, the parameter $q$ in the EoS (\ref{kk1}) cannot be
identified with the cosmological time $t$ anymore. Since we have $f'(q)$
and $f''(q)$ at the de Sitter
point $q=q_n \equiv 2n \pi t_0$, one may imagine that $f(q)$ could be a
constant $f(q)=f(q_n)$ or
\be
\label{pp1}
\rho = - p = \rho_n \equiv \frac{3}{\kappa^2}f(q_n)^2\, .
\ee
Therefore, the fluid can be regarded as a (multiple) cosmological
constant one.
For matter we will now consider dust or cold dark matter and baryonic
matter.
If one of the de Sitter point corresponds to the present universe, the
evolution of the
universe can be approximated by the one corresponding to the $\Lambda$CDM
model, and we have $q_0 = -0.58$ and $j_0 = 1$.
Let $H_\mathrm{present}$ be the present value of the Hubble rate,
$H_\mathrm{present} \sim 70\, \mathrm{km}/\mathrm{s}\cdots \mathrm{Pc}$.
If the present universe corresponds to the de Sitter point, we have
\be
\label{pp2}
\frac{\kappa^2 \rho_n}{3 H_\mathrm{present}^2}
= \frac{f(q_n)^2}{H_\mathrm{present}^2}\sim 0.73\, ,
\ee
which gives a constraint for the parameters of model.
For example, for the model (\ref{p4}), we have
\be
\label{pp3}
\frac{4 \pi^2 n^2 H_0^2 }{H_\mathrm{present}^2}\sim 0.73\, ,
\ee
The deceleration parameter $q_0$ and the jerk parameter $j_0$ will not give
any additional information on the relevant parameters. But if we had more
accurate values of the snap parameter $s_0$ and of
the jerk parameter $l_0$, which are defined as \cite{star}
\be
\label{pp4}
s_0 = \frac{1}{H^4 a}\frac{d^4 a}{dt^4}\, ,\quad
l_0 = \frac{1}{H^5 a}\frac{d^4 a}{dt^5}\, ,
\ee
we could then obtain more constraints on the parameters $t_0$ and $n$.

As the
universe expands, the relative acceleration between two points separated
by a distance $l$ is given by $l \ddot a/a$.
If there is a particle with mass $m$ at each of these points, an observer at
one of the masses will measure an inertial force on the other mass, as
\be
\label{i1}
F_\mathrm{iner}=m l \ddot a/a = m l \left( \dot H + H^2 \right)\, .
\ee
We may estimate the inertial force for the model (\ref{p5}).
At late time, $t\gg t_0$, we find $H\sim \frac{H_0}{t_0}t$ and $H^2 \gg
\left| \dot H \right|$,
therefore,
\be
\label{p14}
F_\mathrm{iner} \sim \frac{m l H_0^2}{t_0^2} t^2 \, .
\ee
If the inertial force becomes larger than the binding energy for bound
states, these bound states are ripped off and destroyed.
This effect explains the disintegration of bound objects in rip universes (Big
Rip, Little Rip or Pseudo-Rip).

\begin{figure}[-!h]
\begin{center}
\includegraphics[angle=0, width=0.7\textwidth]{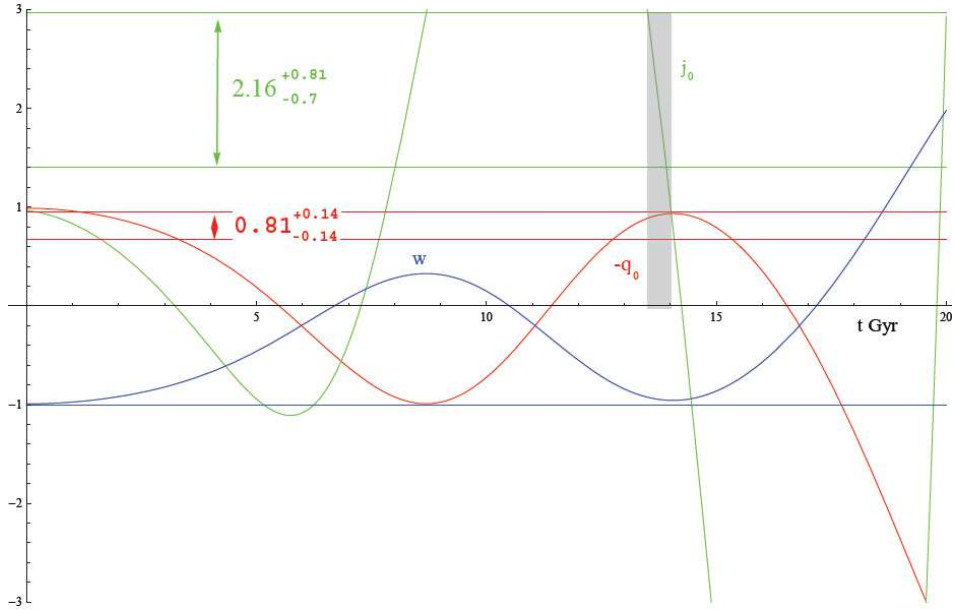}
\caption{\label{Fig5} Plot of $j_0$ (green line), $-q_0$ (red line), and  $w(t)$ (blue
line), for $0\leq t\leq 20$, with $c_1=0.1$, $c=0.447$, $b=2.15$,
$q=1.0445$, $g=3$. The lines of constant time determine the range of
possible values, for the current time, of these quantities. The highlighted time
interval corresponds to the values of $q_0$ and $j_0$ at present.}
\end{center}
\end{figure}

Consider now the more general case
\begin{equation}
\label{ex2}
f(t)=H=\frac{q}{\left(1+c_1 \left(t-b \sin\left(c t\right)\right)\right)^{g}}\, ,
\end{equation}
where $c$, $c_1$, $q$, $b$, and $g$ are constants. Then,
\begin{equation}
\dot{H}=c_1 g q (-1+b c \cos(c t)) \left(1+c_1 t-b\, c_1
\sin \left(c t\right)\right)^{-1-g}\, ,
\end{equation}
and it is easy to see that the time derivative of the Hubble constant will
vanish periodically ($ b \,c \,\cos(c t)=1$). We thus obtain a
model with an effective cosmological constant $p=-\rho$. For the model
(\ref{ex2}), we have
\begin{equation}
w_\mathrm{eff}=-1-\frac{2 c_1 g \left(-1+b\, c \, \cos \left(c t\right)\right)
\left(1+c_1\, t-b\, c_1
\sin \left(c t\right) \right)^{-1+g}}{3 q}\, .
\end{equation}
Note that there is a large arbitrariness in the choice of the constants,
since one can choose them so that the parameters strictly match their current
values (see Fig.~\ref{Fig5}), and one can provide the required stages of the
universe evolution:
Accelerating primordial universe ($-1/3<w<-1$), deceleration of the
universe ($-1/3<w<1/3$), and after that, when $w<-1/3$, the universe turns
into an acceleration phase again. That is,
a transition occurs from the accelerating to the decelerating phase, and
back.

\begin{figure}[-!h]
\begin{center}
\begin{minipage}[h]{0.4\linewidth}
\includegraphics[angle=0, width=1\textwidth]{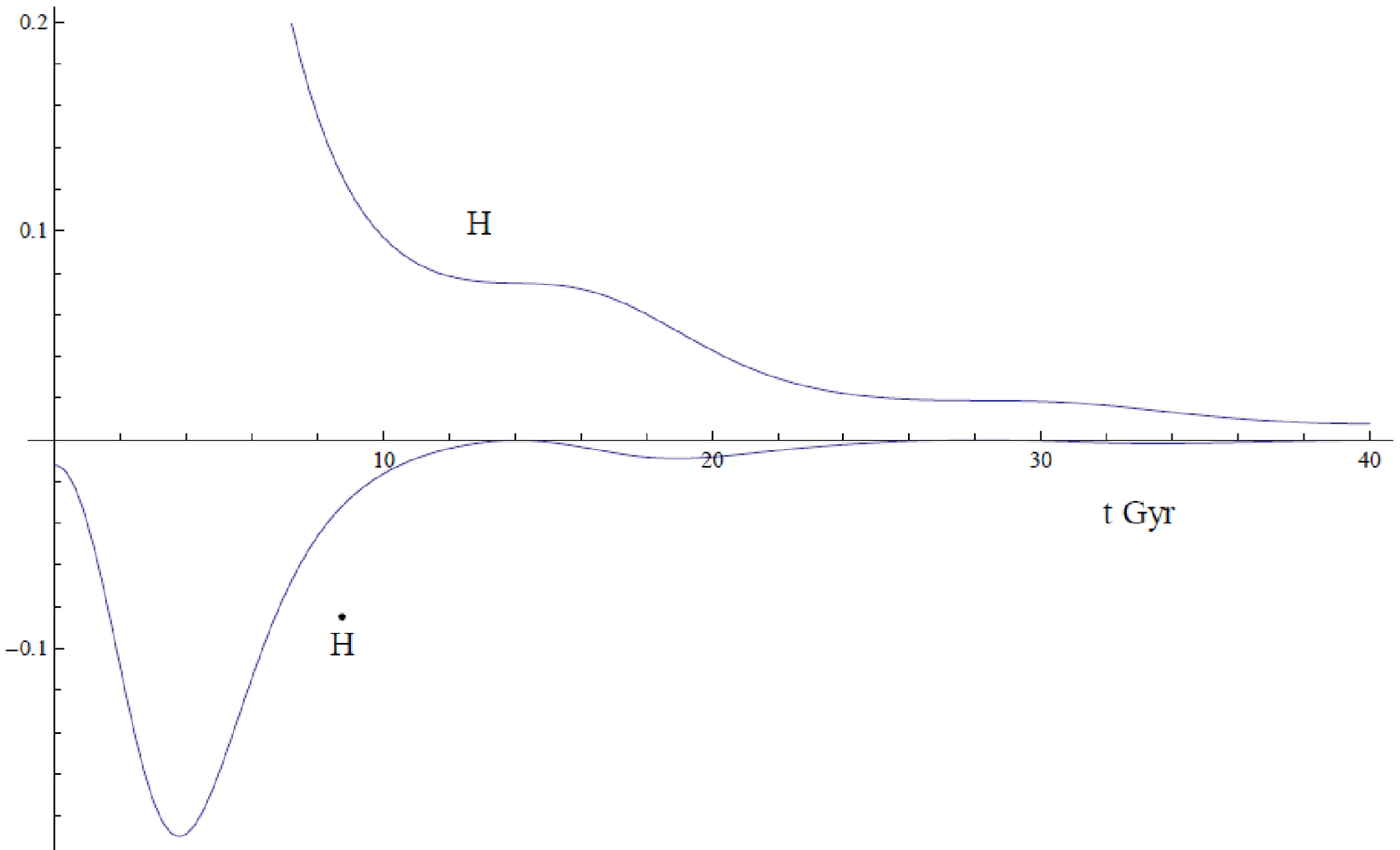}
\caption{\label{Fig6} Plot of $H(t)$ and $\dot{H}(t)$.}
\end{minipage}
\hfill
\begin{minipage}[h]{0.4\linewidth}
\includegraphics[angle=0, width=1\textwidth]{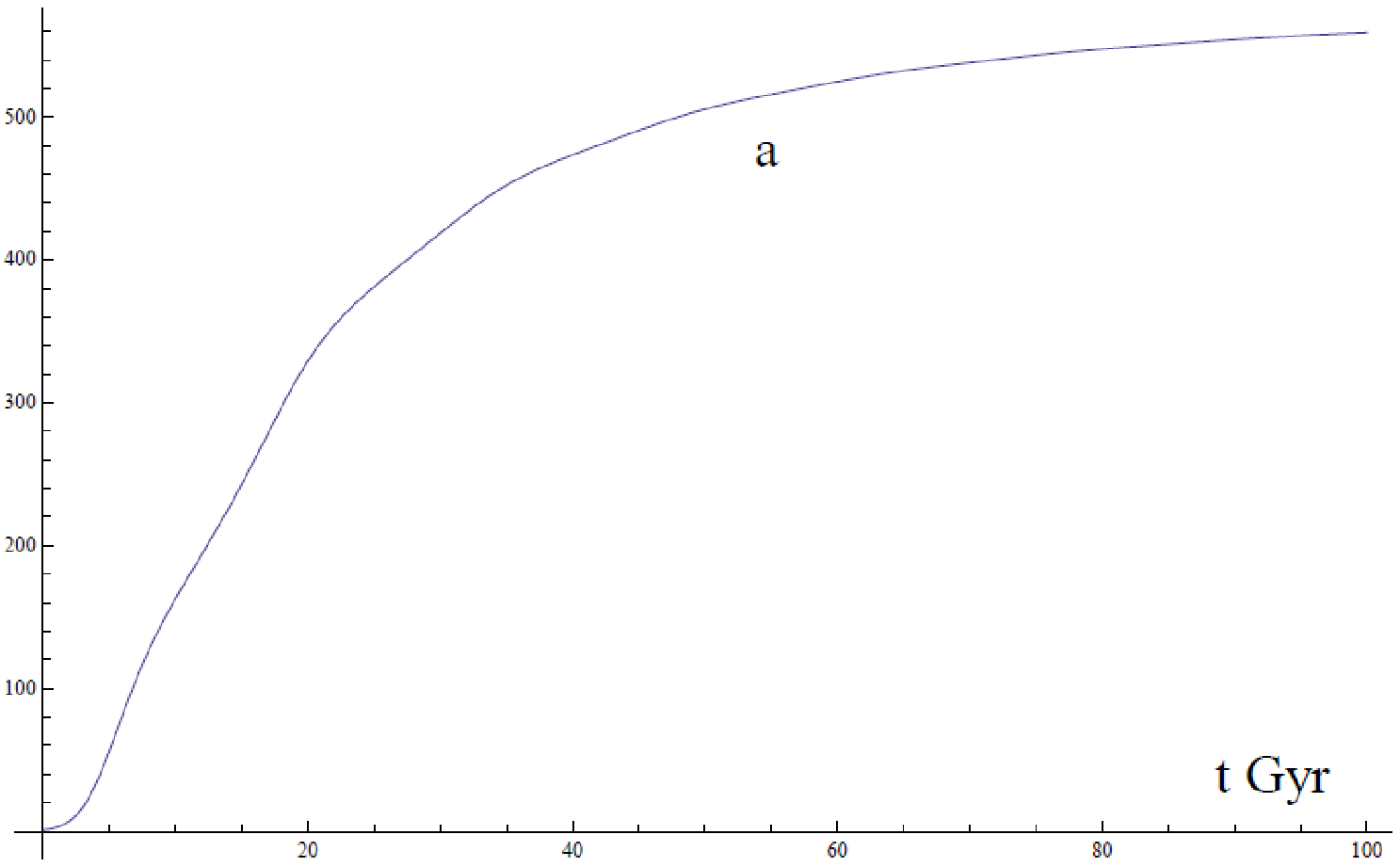}
\caption{\label{Fig7} Plot of $a(t)$.}
\end{minipage}
\end{center}
\end{figure}

For $c_1=0.1$, $c=0.447$, $b=2.15$, $q=1.0445$, $g=3$, and $t=13.6$\,Gyr we
find the following values of the cosmological parameters:
$q_0=-0.902$, $j_0=2.639$, $w=-0.935$, and $H_0=0.0752$\,Gyr$^{-1}$.
All these values correspond to the measured values at the current time
($t=13.6$\,Gyr).
Thus, with the pass of time both the cosmological constant and its
derivative, and with them the energy density and pressure too, will tend
to zero (see Figs.~\ref{Fig6} and \ref{Fig7}).
One can easily see that $H\to 0$ for
$t->\infty$ and, hence, those are pseudo-Rip models.

\begin{figure}[-!h]
\begin{center}
\begin{minipage}[h]{0.4\linewidth}
\includegraphics[angle=0, width=1\textwidth]{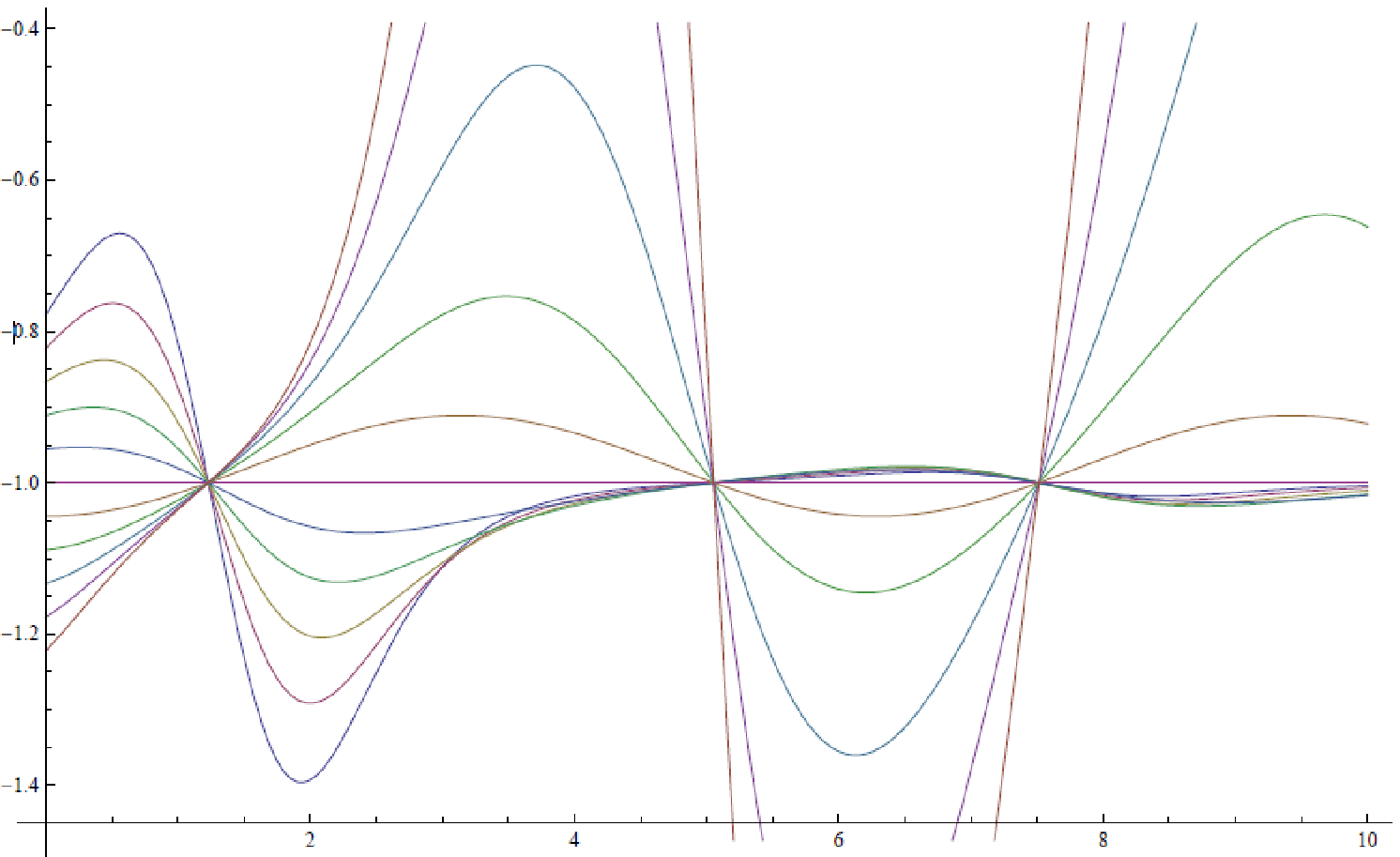}
\caption {\label{Fig8} Plot of $w(t)$ ($0\leq t \leq 10$), for $c_1=1$,
$c=0.1$, $b=3$, $q=3$, and $-5< g<5$.}
\end{minipage}
\hfill
\begin{minipage}[h]{0.4\linewidth}
\includegraphics[angle=0, width=1\textwidth]{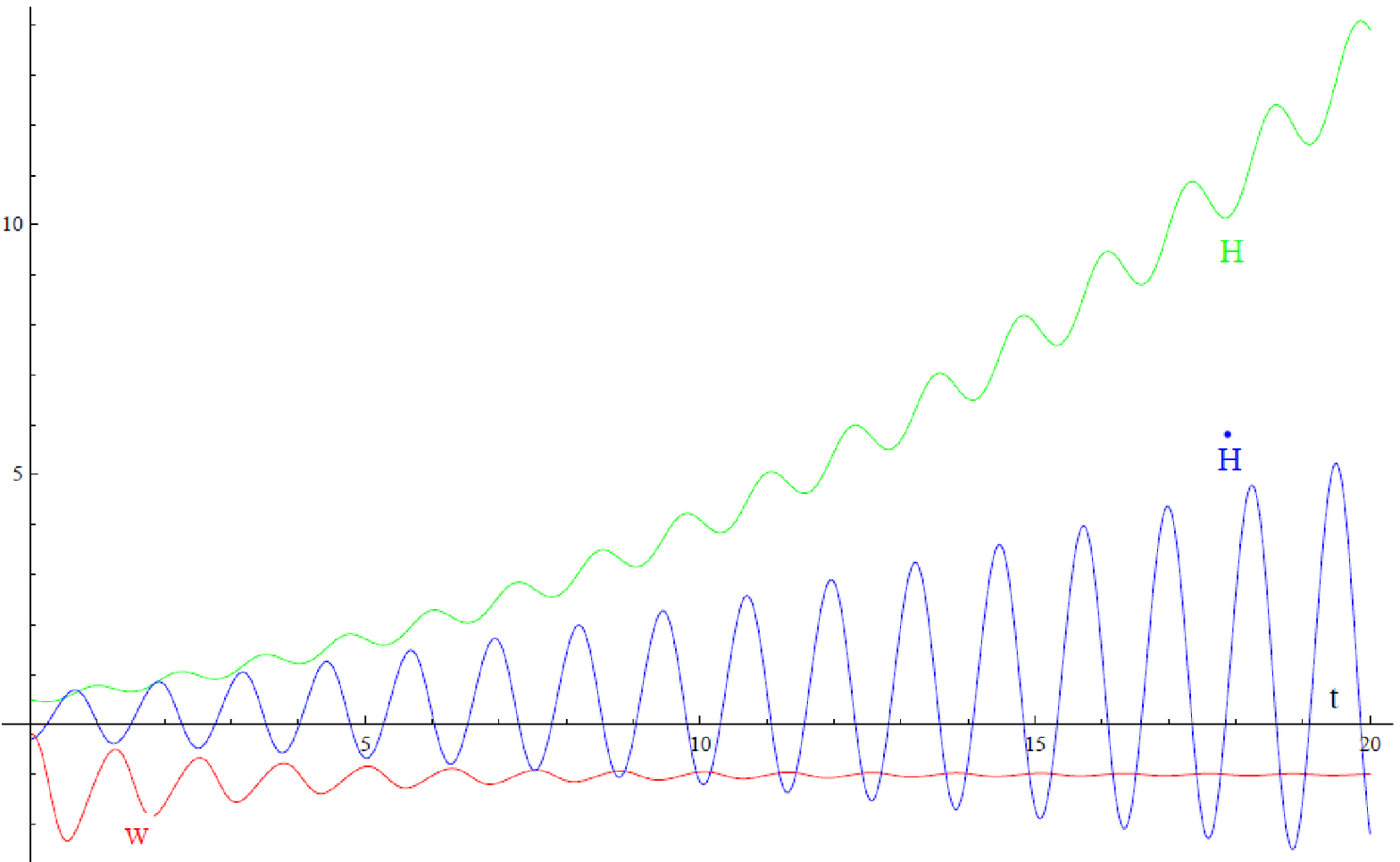}
\caption{\label{Fig9} Plot of $H$ (green line), $w$ (red line), and
$\dot{H}(t)$ (blue line), for $c_1=0.1$, $c=5$, $b=0.6$, $q=0.5$, and $g=-3$.}
\end{minipage}
\end{center}
\end{figure}

By selecting different values of the constants one can obtain different
behaviors for the EoS parameter (see Fig.~\ref{Fig8}):
\begin{enumerate}
\item  For the earlier values of time one gets accelerated
expansion, then the expansion slows down, and later the acceleration
will start again.
\item The oscillation $w$ can acquire values around minus one (see
Fig.~\ref{Fig9}). This case corresponds to the Little Rip model ($H\to\infty$ for
$t\to \infty$).
\end{enumerate}

\begin{figure}[-!h]
\begin{center}
\includegraphics[angle=0, width=0.7\textwidth]{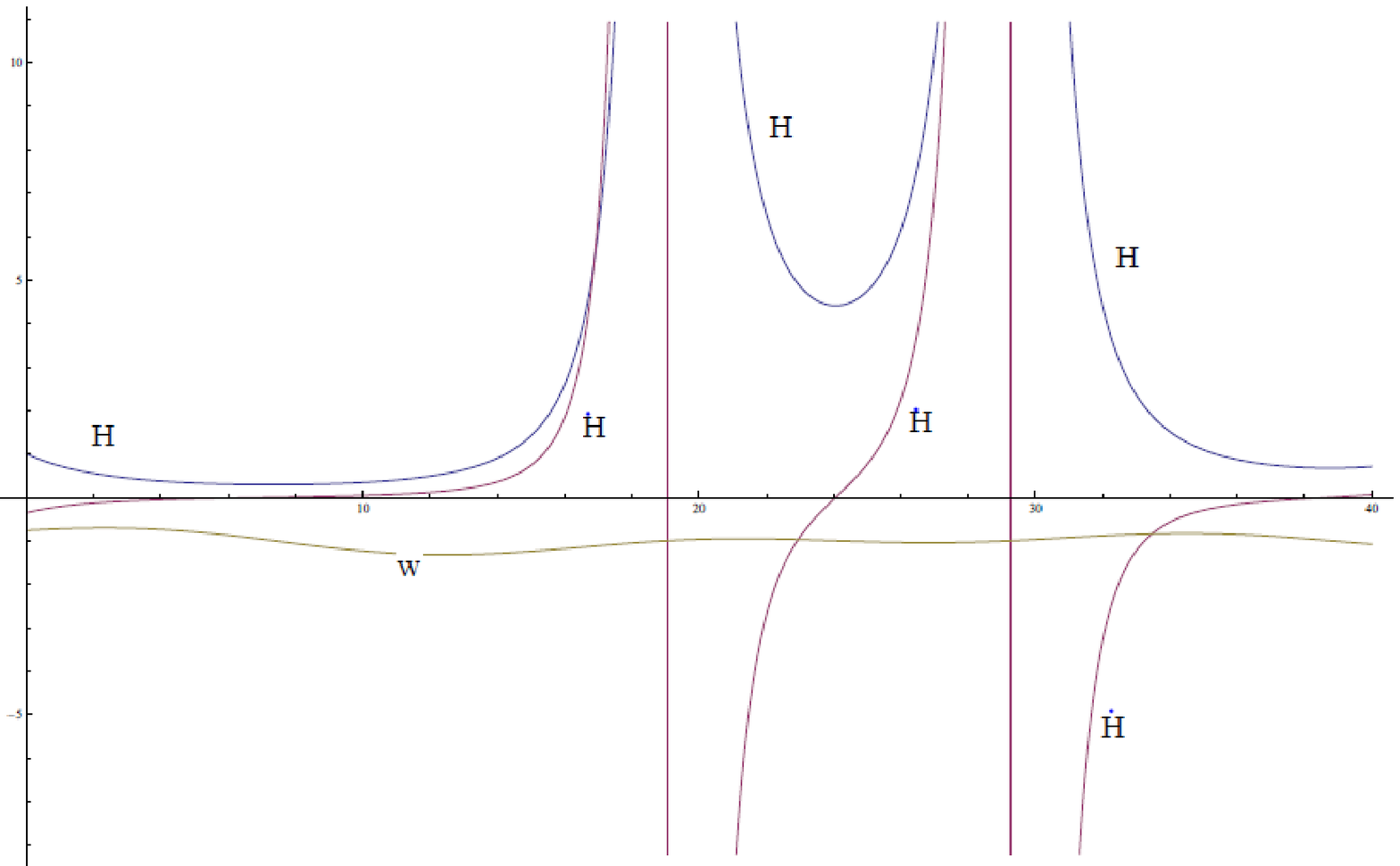}
\caption{\label{Fig10} Typical behavior the model considered, for singularities of Big
Rip type. In the present model, $g=2$.}
\end{center}
\end{figure}

If $\gamma$ is positive and the parameter $c_1$ is negative, then we get a
singularity in the future. This situation was already discussed above.
By choosing proper values of the constants, different future singularities can
be obtained as, for instance, the one depicted in Fig.~\ref{Fig10}.

It can be seen that a model of this kind leads to different types of
evolution of the universe. First, one can build a model that will
consistently describe all the stages in the universe evolution:
accelerated expansion, slowing down to $w = 1/3$, and accelerated
expansion again, while for $t\to\infty$ the Hubble constant and its
derivative tend to zero.
Second, one can adjust for the right behavior of the model in the far
future: The universe turns to be de Sitter or exhibits one of the four
types of singularities. Moreover, almost always is it possible to choose the
parameters so that they match the observed values. This is not difficult
to do by assuming that at present the universe is in a phase
corresponding to an effective cosmological constant. In addition, these
models exhibiting multiple cosmological constants may
coexist simultaneously, which definitely shows an analogy with the
cosmological landscape picture.

\section{Cosmological reconstruction by one scalar model}

We now construct scalar field models realizing the cosmological fluids given in the
previous sections.
The formulation is based on \cite{grg} and we shall start with the following action:
\be
\label{ma7}
S=\int d^4 x \sqrt{-g}\left\{
\frac{1}{2\kappa^2}R - \frac{1}{2}\omega(\phi)\partial_\mu \phi
\partial^\mu\phi - V(\phi) + L_\mathrm{matter} \right\}\, .
\ee
Here, $\omega(\phi)$ and $V(\phi)$ are functions of the scalar field $\phi$.
The function $\omega(\phi)$ is not relevant, as it can be absorbed into the
redefinition of the scalar field $\phi$, as follows,
\be
\label{ma13}
\varphi \equiv \int^\phi d\phi \sqrt{\left|\omega(\phi)\right|} \, .
\ee
The kinetic term of the scalar field in the action (\ref{ma7}) has the
following form:
\be
\label{ma13b}
 - \omega(\phi) \partial_\mu \phi \partial^\mu\phi
= \left\{ \begin{array}{ll}
 - \partial_\mu \varphi \partial^\mu\varphi &
\mbox{when $\omega(\phi) > 0$} \\
\partial_\mu \varphi \partial^\mu\varphi & \mbox{when $\omega(\phi) < 0$}
\end{array} \right. \, .
\ee
The case $\omega(\phi) > 0$ corresponds to quintessence or a
non-phantom scalar field, and the case of $\omega(\phi) < 0$ corresponds
to a phantom scalar.
Although $\omega(\phi)$ can be absorbed into the redefinition of the
scalar field, we keep $\omega(\phi)$ since the transition between the
quintessence and phantom cases can be best described by the change of sign
of $\omega(\phi)$.

In order to consider and explain the cosmological reconstruction in terms of
one scalar model, we rewrite the FRW equation as follows:
\be
\label{ma9}
\omega(\phi) {\dot \phi}^2 = - \frac{2}{\kappa^2}\dot H\, ,\quad
V(\phi)=\frac{1}{\kappa^2}\left(3H^2 + \dot H\right)\, .
\ee
Assuming $\omega(\phi)$ and $V(\phi)$ are given by a single function
$f(\phi)$, as
\be
\label{ma10}
\omega(\phi)=- \frac{2}{\kappa^2}f'(\phi)\, ,\quad
V(\phi)=\frac{1}{\kappa^2}\left(3f(\phi)^2 + f'(\phi)\right)\, ,
\ee
we find that the exact solution of the FLRW equations
(when we neglect the contribution from  matter)
has the following form:
\be
\label{ma11}
\phi=t\, ,\quad H=f(t)\, .
\ee
It can be confirmed that the equation given by the variation over $\phi$,
\be
\label{ma12}
0=\omega(\phi)\ddot \phi + \frac{1}{2}\omega'(\phi){\dot\phi}^2
+ 3H\omega(\phi)\dot\phi + V'(\phi)\, ,
\ee
is also satisfied by the solution (\ref{ma11}).
Then, the arbitrary universe evolution expressed by $H=f(t)$ can be
realized by an appropriate choice of $\omega(\phi)$ and $V(\phi)$.
In other words, defining the particular type of universe evolution,
the corresponding scalar-Einstein gravity can be found.

For example, for the model  (\ref{MLa1}), we get
\bea
\label{MA1}
\omega(\phi) &=& \frac{2 H_0 g}{\kappa^2} \left(1 - \cos \omega \phi\right)
\e^{ - g \left( 1 - \frac{1}{\omega} \sin \omega \phi \right)}\,
f'(\phi)\, ,\nonumber \\
V(\phi) &=& \frac{1}{\kappa^2}\left(3
H_0^2 \e^{ - 2 g \left(t - \frac{1}{\omega} \sin \omega \phi \right)}
  - H_0 g \left(1 - \cos \omega \phi\right)
\e^{ - g \left( 1 - \frac{1}{\omega} \sin \omega \phi \right)}\right)\, ,
\eea
and, for the model (\ref{p4}),
\bea
\label{MA2}
\omega(\phi) &=& - \frac{2H_0}{\kappa^2 t_0} \left( 1 - \cos \frac{\phi}{t_0}
\right)\, ,
\nonumber \\
V(\phi) &=& \frac{1}{\kappa^2}\left(3
H_0^2 \left( \frac{t}{t_0} - \sin \frac{t}{t_0} \right)^2
+ \frac{H_0}{t_0} \left( 1 - \cos \frac{t}{t_0} \right) \right)\, ,
\eea
In the same way we can obtain the scalar theory corresponding to any of
the other models described by a dark fluid with an EoS of the types
considered above.

\section{Conclusions}

We have built in this paper several dark energy models, with a
time-dependent equation of state, which can be viewed as simple classical
analogs of the string landscape. The possible (simultaneous) existence of
several cosmological constants can be interpreted as the possible presence
of several vacuum states one has to choose from, what could bring into
play Casimir effect considerations. Their simultaneous occurrence may
indicate a future transition to a $\Lambda$CDM epoch with a different
value for the effective cosmological constant.

It is very interesting to realize that the freedom we actually have in
those models allows us in many cases, on top of providing a reasonable
description of the different epochs of the universe evolution, to also
adjust for their right behavior in the far future: the universe turns to
be (asymptotically) de Sitter or exhibits one of the four types of
finite-time future singularities or shows a Little Rip behavior.
Moreover, up to some exceptions, it is
possible to choose the parameters so that they match the astronomical
data providing a very realistic description of $\Lambda$CDM cosmology.
This is not difficult to do by assuming that, at the current moment of
its evolution,
the universe is in a phase corresponding to a given effective cosmological
constant. Remarkably, the different models, which correspond to different
cosmological constants, could coexist at the same moment, which definitely
hints to an intriguing classical analogy with the cosmological landscape
picture. From another viewpoint, the rich structure of the cosmological (singular)
behavior of the models under discussion indicates that maybe similar phenomena
could be typical in the string landscape.

The important lesson to be taken from current investigation is that, even if our
current universe may look as the one described with the help of an
effective cosmological constant, its finite-time future may be singular, so
that its evolution might effectively end up. This opens the problem of the
interpretation of the more precise observational data to come,
which should be tailored with the specific purpose to understand what future
is favored by the cosmological bounds this data will undoubtedly impose.

\section*{Acknowledgments.}

EE's research was performed while on leave of absence at Dartmouth College,
NH, USA, supported by MICINN (Spain), contract PR2011-0128. EE and SDO
have been partly supported by MICINN (Spain),  projects
FIS2006-02842 and FIS2010-15640, by the CPAN
Consolider Ingenio Project, and by AGAUR (Generalitat de Ca\-ta\-lu\-nya),
contract 2009SGR-994.
ANM was supported by the ESF Programme ``New Trends and Applications of
the Casimir Effect'' (Short Visit 4687). VVO and ANM are
grateful to the  LRSS project No 224.2012.2.
SN is supported in part by Global COE Program of Nagoya University (G07)
provided by the Ministry of Education, Culture, Sports, Science \&
Technology and by the JSPS Grant-in-Aid for Scientific Research (S) \# 22224003
and (C) \# 23540296.

\end{document}